\documentclass[preprint2]{aastex} 
\usepackage{psfig}

\begin{document}

\title{A 695-Hz quasi-periodic oscillation in the
low-mass X-ray binary EXO 0748--676} 

\author{Jeroen Homan \& Michiel
van der Klis} 

\affil{Astronomical Institute 'Anton Pannekoek',
University of Amsterdam, and Center for High Energy Astrophysics}
\affil{ Kruislaan 403, 1098 SJ, Amsterdam, The Netherlands}

\begin{abstract}
We report the discovery of a 695-Hz quasi-periodic oscillation (QPO)
in data taken with the {\it Rossi X-ray Timing Explorer} of the
low-mass X-ray binary (LMXB) EXO 0748--676. This makes EXO 0748--676
the second dipping LMXB, after 4U 1915--05, that shows kHz
QPOs. Comparison with other sources suggests that the QPO corresponds
to the lower frequency peak of the kHz QPO pair often observed in
other LMXBs. The QPO was found in the only observation done during an
outburst of the source in early 1996. This observation is also the
only one in which the $\sim$1 Hz QPO recently found in EXO 0748--676
is not present.
\end{abstract}

\keywords{accretion, accretion disks -- stars: individual (EXO
0748--676) -- X-rays: stars}

\section{Introduction}
High frequency (kHz) quasi-periodic oscillations (QPOs) have been
found in many neutron-star low-mass X-ray binaries (see van der Klis
\markcite{va2000}2000 for a recent review). They are observed in the
300--1300 Hz range, and are often found in pairs with a nearly
constant frequency separation of $\sim$250--350 Hz. In addition to kHz
QPOs, some sources have shown slightly drifting oscillations in the
330--590 Hz range, during type-I X-ray bursts (Strohmayer, Swank, \&
Zhang \markcite{stswzh98}1998).

In this paper we present our search for both kHz QPOs and burst
oscillations in the low-mass X-ray binary EXO 0748--676. This source
shows periodic (P=3.82 hr) eclipses, irregular intensity dips, and
type-I X-ray bursts (Parmar et al. \markcite{pawhgi1986}1986). From
the eclipse duration a source inclination of 75$^\circ$ to 82$^\circ$
was derived (Parmar et al. \markcite{pawhgi1986}1986). Based on its
bursting behavior (e.g. burst rate and peak flux vs. persistent flux;
see Gottwald et al. \markcite{gohapa1986}1986) EXO 0748--676 may be
a member of the atoll class (Hasinger \& van der Klis
\markcite{hava1989}1989) of the neutron-star low-mass X-ray
binaries. Recently, a variable 0.58--2.44 Hz QPO was found by Homan et
al. (1999). This QPO was found in all observations, except in the only
observation during an outburst of the source (early 1996) observed
with the {\it Rossi X-ray Timing Explorer} (RXTE), and is probably
caused by an orbiting structure in the accretion disk, which modulates
the radiation of the central source (Jonker et
al. \markcite{jowiva1999}1999; Homan et
al. \markcite{hojowi1999}1999).

\section{Observations and Analysis}

\begin{figure}[t]
\psfig{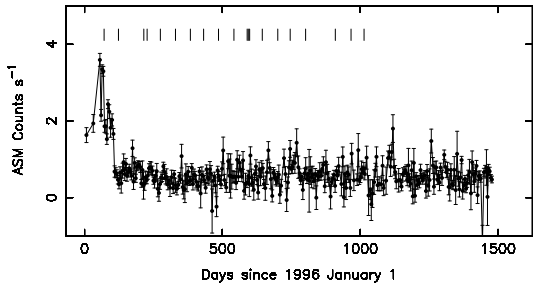}
\caption{The four-day average RXTE/ASM light curve of EXO
0748--676. The short vertical lines depict the times of the RXTE/PCA
observations analyzed in this paper. \label{asm_fig}}
\end{figure}

The data used in this paper were obtained with the Proportional
Counter Array (PCA; Jahoda et al. \markcite{jaswgi1996}1996) onboard
RXTE (Bradt, Rothschild, \& Swank \markcite{brrosw1993}1993), between
March 12 1996 and October 11 1998.  Most PCA observations were done in
sets of five or six $\sim$2 ks segments that were centered around
successive eclipses. Data of these five or six segments were taken
together and treated as single observations, which resulted in a total
of 20 observations. The times of the observations are indicated in
Figure \ref{asm_fig}, which also shows the RXTE All Sky Monitor (ASM)
light curve of EXO 0748--676. As can be seen, only the first
observation was done during the early 1996 outburst of the source.

The PCA data were obtained in several different modes. The Standard 1
and 2 modes, which were always active, had 1/8 s time resolution in
one energy channel (1--60 keV, representing the full energy range
covered by the PCA), and 16 s time resolution in 129 energy bands
(1--60 keV), respectively. In addition to the two Standard modes,
another mode was always active which had a time resolution better than
1/8192 s in at least 32 energy channels (1--60 keV).

The Standard 2 data were used to produce light curves and hard color
curves. The hard color was defined as the ratio of the count rates in
the 6.3--11.7 keV and 5.2--6.3 keV bands; the light curves were
produced in the 1.6--14.4 keV band, i.e. the energy band in which,
during the first observation, the count rate spectrum of the source
exceeded that of the background. Using the high time resolution data,
0.0625--2048 Hz power spectra were created in several energy bands to
search for kHz QPOs. The power spectra were selected on time, count
rate, or hardness, before they were averaged. The average power
spectrum was rms renormalized (van der Klis \markcite{va1995}1995),
and fitted (in the 100-1500 Hz range) with a constant for the Poisson
level, and a Lorentzian for any QPO (only one QPO was found). Errors
on the fit parameters were determined using $\Delta\chi^2=1$
(1$\sigma$, single parameter). As significance of each power spectral
feature we quote the inverse relative error on the integrated power of
each feature, as measured from the power spectrum. Note that the
inverse relative error on the fractional amplitude (the parameter we
give) is larger than the true significance by a factor 2.  The energy
dependence of the QPO was determined by fixing the frequency and width
of the QPO to their values obtained in the band where the QPO was most
significant (6.6--18.7 keV). Upper limits on kHz QPOs were determined
in the 100--1500 Hz range by fixing the width of the QPO to 10, 20, 50
or 100 Hz, and using $\Delta\chi^2=2.71$ (95\% confidence). Upper
limits were only determined in the total (1--60 keV) band, and in the
band where the detected QPO was most significant. To determine upper
limits for oscillations in the type-I X-ray bursts, 2--1024 Hz power
spectra were created. The width of the QPO was fixed to 2 Hz.

\section{Results}

\begin{figure}[t]
\psfig{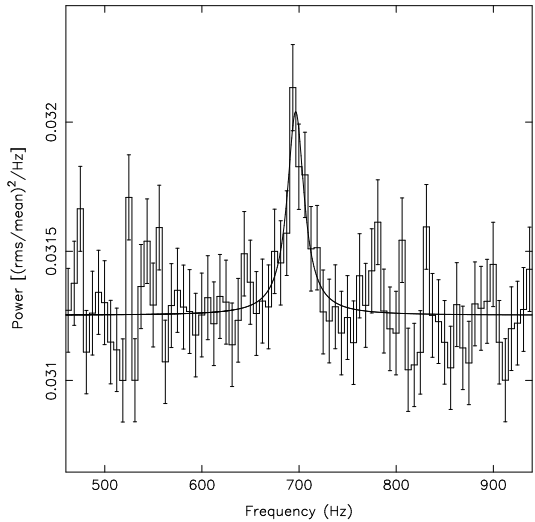}
\caption{The 6.6--18.7 keV power spectrum of EXO 0748--676
for the 1996 March 12 observation. The solid line shows the best fit,
with a Lorentzian at 695 Hz.\label{kHz_fig}}
\end{figure}

\begin{deluxetable}{ccccc}
\tablecolumns{5}

\tablehead{ \colhead{Selection} & \colhead{Values} & \colhead{Frequency (Hz)} &
\colhead{FWHM (Hz)} & \colhead{rms amplitude (\%)}}

\tablecaption{QPO parameters in the 6.6-18.7 keV band for data selected on time (since start of
observation), 1.6--14.4 keV count rate, or hard color, for the 1996
March 12 observation.\label{selection_tab}}
\startdata
Time (s)    & & & & \\
            & 0--2000     & 693.6$^{+0.8}_{-1.2}$ & 5$\pm$3 & 11.5$^{+1.8}_{-1.6}$ \\
            & 2000--4200  & 696.1$^{+1.9}_{-1.7}$ & 15$^{+6}_{-4}$ & 17$\pm$2 \\
            & 6000--7000  & 698$\pm$3 & 18$^{+7}_{-5}$ & 19$\pm$3 \\
            & 7000--7900  & 708$^{+5}_{-3}$ & 13$^{+15}_{-10}$ & 18$\pm$5\\

Count rate (cts/s) & & & & \\
                   & $<$285      & 692$^{+2}_{-3}$ & 15$^{+6}_{-4}$ & 15$\pm$2 \\
                   & $>$285      & 695.0$^{+1.1}_{-0.9}$ & 9$\pm$3 & 14.2$^{+1.9}_{-1.4}$ \\
Hard Color  &       & & & \\  
            & $<$1.2      & 694$^{+2}_{-3}$ & 15$^{+10}_{-6}$ & 14.4$^{+2.8}_{-1.7}$ \\
            & $>$1.2      & 694.2$^{+0.9}_{-0.8}$ & 9$^{+3}_{-2}$ &  14.8$^{+1.6}_{-1.4}$\\

\enddata
\end{deluxetable}

\begin{figure}[t]
\psfig{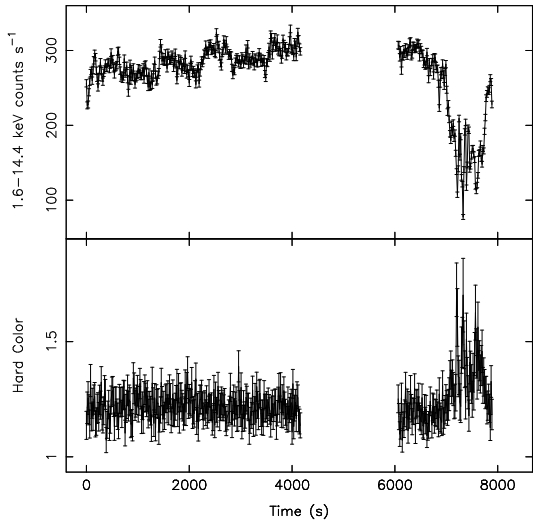}
\caption{The 1.6--14.4 keV light curve (a) and
hard color curve (b) of the 1996 March 12 observation. For definition
of hard color, see text. The data gap between 4000 and 6000 s is due
to a passage of RXTE through the south Atlantic
anomaly.\label{lc-hc_fig}}
\end{figure}

\begin{figure}[t]
\psfig{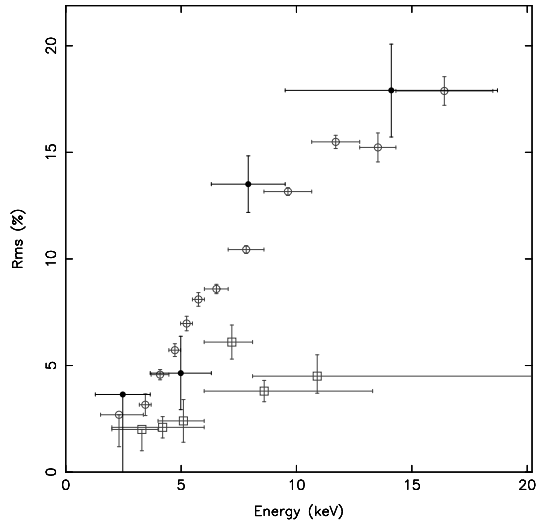}
\caption{Energy dependence of the 695 Hz QPO in
the 1996 March 12 observation (bullets). For comparison the energy
dependencies of the lower peak (open circles) and upper peak (squares)
of 4U 1608--52 are shown.\label{energy_fig}}
\end{figure}

Our search for kHz QPOs in EXO 0748--676 resulted in only one
significant detection: a $\sim$695 Hz QPO was found in the 1996 March
12 observation, the only observation done during the early 1996
outburst. In the 6.6--18.7 keV band, where it was found to be most
significant, it had a frequency of 695.0$\pm$1.2 Hz, a
full-width-at-half-maximum (FWHM) of 14$^{+4}_{-3}$ Hz, and an rms
amplitude of 15.2$^{+1.4}_{-1.3}$\% (6.0$\sigma$, single trial). The
6.6--18.7 keV power spectrum of the 1996 March 12 observation is shown
in Figure \ref{kHz_fig}. In the 1--60 keV band the QPO had a frequency
of 693.0$^{+0.5}_{-0.8}$ Hz, a FWHM of 3.9$^{+3.2}_{-0.4}$ Hz, and an
rms amplitude of 5.4$^{+1.0}_{-0.7}$\% (3.7$\sigma$, single trial).
The 1.6--14.4 keV light curve and the hard color curve of the 1996
March 12 observation are shown in Figure \ref{lc-hc_fig}. In order to
examine the variability of the QPO in the 6.6--18.7 keV band,
selections were made on time, 1.6--14.4 keV count rate, and hard
color. The selections for count rate and hard color were performed
only for the data before the dip ($t<7000\,s$). The results for the
different selections are shown in Table \ref{selection_tab}. The QPO
frequency increased slightly with time and perhaps count rate, but it
did not depend on hard color. Note that the detection of the QPO in
the dip, at 708 Hz, is only at a 1.8$\sigma$ significance level (in
the 6.6--18.7 keV band). We made some subselections on the last time
interval (which includes the dip) to test whether the QPO amplitude
varied with decreasing count rate; only upper limits could be
determined for those subselections, with values higher than obtained
for the whole selection. The energy dependence of the QPO, in the part
before the dip, was determined using four energy bands and is shown in
Figure \ref{energy_fig}. The QPO energy spectrum is rather steep, with
an upper limit of 3.6\% in the lowest energy band and an rms amplitude
of 18\% in the highest energy band. For reasons of comparison we also
plotted the energy dependence of the lower and upper kHz QPOs of 4U
1608--52 (Berger et al. \markcite{bevava1996}1996; M\'endez et
al. \markcite{mevava1998}1998; M\'endez et al.  \markcite{me2000}2000)

Upper limits (6.6--18.7 keV) on any second kHz QPO (as often observed
in other low-mass X-ray binaries) were determined in the 100--1500 Hz
range. They were 9.5\%, 10.5\%, 14.3\%, and 16.7\% rms, for fixed
widths of 10 Hz, 20 Hz, 50 Hz, and 100 Hz, respectively. Since the
frequency of the QPO varied little, the ``shift and add'' method
(M\'endez et al. \markcite{mevava1998}1998) could not usefully be
applied. No $\sim$1 Hz QPO was found either, as was already reported
by Homan et al. (\markcite{hojowi1999}1999). Upper limits (1--60 keV)
are $\sim$7\% rms in the 0.001--1 Hz range, $\sim$2\% rms in the 1--10
Hz range, and $\sim$4\% rms in the 10--50 Hz range.

For the other observations only upper limits to the presence of a kHz
QPO could be determined. This was done in the 1--60 keV and 6.6--18.7
keV bands. The upper limits are given in Table \ref{upper_tab}, for
four different fixed widths.  Most of the upper limits are comparable
to or larger than the values found for the QPO in the 1996 March 12
observation, and therefore not very constraining. Note that the
count rate in these observations was a factor 2 to 3 smaller than that
during the 1996 March 12 observation, resulting in a greatly reduced
sensitivity. A similar kHz QPO as the one seen there would, when
present, be significant only at a $\sim1\sigma$ level. In all these
observations a $\sim$1 Hz QPO was found, with a frequency between
$\sim$0.4 and $\sim$3 Hz.

Ten type-I X-ray burst were observed, and they were examined for the
presence of burst oscillations. This was done in the 100--1000 Hz
frequency range, for a fixed width of 2 Hz. None were found, with
upper limits during the rise of the burst between 4\% and 11\% rms in
the 1--60 keV band, and between 6\% and 14\% rms in the 6.6--18.7 keV
band. This is well below the amplitudes of burst oscillations observed
in some other sources (e.g. Strohmayer et
al. \markcite{stzhsw1998}(1998); see also van der Klis
\markcite{va2000}2000 and references therein).  It should be noted that
the rise times of the bursts were rather long, between 2 and 12
s. This could be an indication for the presence of a scattering medium
surrounding the neutron star, which might wash out the rapid burst
oscillations. The $\sim$1 Hz QPO was observed in all the bursts (see
also Homan et al. \markcite{hojowi1999}1999).

\section{Discussion}

\begin{deluxetable}{ccccc}
\tablecolumns{5}

\tablehead{ \colhead{Energy band (keV)} & \colhead{rms amp. (\%)} & \colhead{rms amp. (\%)} &
\colhead{rms amp. (\%)} & \colhead{rms amp. (\%)}}

\tablecaption{95\% confidence upper limits for kHz QPOs in
non-outburst power spectra, in two energy bands and for four different
fixed FWHM. Ranges represent the lowest and highest upper limits
measured among all observations.\label{upper_tab}}

\startdata
          & FWHM=10 Hz & FWHM=20   & FWHM=50    & FWHM=100 \\ 
1--60     & 6.4--13.9  & 9.4--14.9 & 10.7--17.6 & 11.6--21.7  \\
6.6--18.7 & 6.3--13.6  & 6.6--16.3 & 8.2--19.5  & 9.2--23.0  \\

\enddata
\end{deluxetable}

The properties of the 695 Hz QPO are similar to those of the kHz QPOs
in atoll sources; the QPO is relatively narrow (5--18 Hz) and
has an rms amplitude of $\sim$6.5\% (1--60 keV, outside the
dip). Since only a single peak is observed, we can not tell whether it
corresponds to the lower or the upper peak of a kHz QPO pair. However,
comparison with kHz QPOs in atoll sources (see van der Klis
\markcite{va2000}2000) suggests that the observed QPO is the lower QPO
of a kHz pair, for the following reasons: (1) of the 11 kHz QPO pairs
found in atoll sources, 8 have ranges of lower peak frequencies that
include 695 Hz, which is the case for only 3 of the upper peaks. (2)
The upper peaks in atoll sources have widths in the 50--200 Hz range,
although occasionally peaks with widths of only 10 to 20 Hz have been
observed. On the other hand, the 4--18 Hz width we find is much more
common for lower peaks. (3) When comparing the energy dependence of
the QPO with that of the two kHz peaks in 4U 1608--52, which have
rather different energy dependencies (Berger et
al. \markcite{bevava1996}1996; M\'endez et
al. \markcite{mevava1998}1998; M\'endez et al. \markcite{me2000}2000),
we find that it was very similar (i.e. steep) to that of the lower
peak (see Figure \ref{energy_fig}). Hence three of the QPO properties
hint towards the QPO being the lower peak.

The properties of the QPO varied on a time scale of a few $10^3$ s, as
can be seen from Table \ref{selection_tab}. Comparing the first time
selection with the second, one can see that a relatively small
frequency change is accompanied by a factor 3 (2$\sigma$) increase in
width, and an almost 50\% (2$\sigma$) increase in fractional rms
amplitude.

The other source in which only a single kHz QPO has been observed is
XTE J1723--376 (Marshall \& Markwardt
\markcite{mama1999}1999). However, most sources in which kHz QPO pairs
have been found, have at times also shown single kHz QPOs. The fact
that EXO 0748--676 and XTE J1723--376 have only shown single kHz QPOs
is therefore most likely  a matter of a small amount of data and
coincidence. 

With $i=75^\circ-82^\circ$ EXO 0748--676 is probably the source with
the highest inclination angle of the $\sim$20 sources that have shown kHz
QPOs. Twin kHz QPOs were already found in 4U 1915-05 (Barret et
al. \markcite{baolbo1997}1997, \markcite{baolbo2000}2000; Boirin et
al. \markcite{bobaol2000}2000), a source that also shows dips (but no
eclipses, which for a similar mass ratio would imply a lower
inclination than EXO 0748--676). The fact that kHz QPOs are found over
a large range of inclinations means that the radiation modulated by
the kHz QPO mechanism should to a large extent be isotropic. The kHz
QPO was detected during the dip, but at a significance of only
1.8$\sigma$. This means that with $\sim$90\% confidence we can say
that either the source producing the kHz QPO was not fully covered by
the dipping material, or that a considerable amount of the modulated
radiation went through the dipping material unperturbed, indicating
that it has a scattering optical depth of at most a few. Also, the
fact that the rms amplitude changes only a little in the dip suggests
that the kHz QPO and the bulk of the flux are produced at the same
site.

The outburst of EXO 0748--676 in early 1996 (see Fig. \ref{asm_fig})
may have been a transition from the island state to the banana state,
and back, as is common for atoll sources. In addition to the increase
in count rate, there are several power spectral properties that seem
to confirm this idea: (1) The strength of 0.1--1.0 Hz noise during the
outburst was lower than in the non-outburst observations (see Homan et
al. 1999). Most atoll sources show a decrease of the noise strength
when they move from the island to the banana state (Hasinger \& van
der Klis \markcite{hava1989}1989). (2) The $\sim$1 Hz QPO was not
observed during the only outburst observation. In 4U 1746--37, one of
the other two sources were a similar $\sim$1 Hz QPO was found, the QPO
was observed only in the island state, and not in the banana state
(Jonker et al. \markcite{jovaho}2000). (3) Although there are a few
exceptions, in most atoll sources kHz QPOs are found only in the lower
banana state (van der Klis \markcite{va2000}2000).

We find the kHz QPO in the only observation where the $\sim$1 Hz QPO
was absent.  The $\sim$1 Hz QPO is thought to be due to obscuration of
the central source by an orbiting structure in the accretion disk at a
distance of $\sim$1000 km from the central source (Jonker et
al. \markcite{jowiva1999}1999; Homan et
al. \markcite{hojowi1999}1999). It is interesting to see that in two
of the sources were the $\sim$1 Hz QPOs are found, they are not
observed in the banana state, indicating a change in the accretion
disk geometry (at least in the area where the $\sim$1 Hz QPO is
formed). This, together with the fact that in most atoll sources kHz
QPO are only found in the banana state, suggests that changes in the
accretion disk geometry (at $\sim$1000 km from the central source) may
affect the production of kHz QPOs close to the central source.

\acknowledgments The authors would like to thank Mariano M\'endez and
Peter Jonker for their help and stimulating discussions. This work was
supported by NWO Spinoza grant 08-0 to E.P.J. van den Heuvel, by the
Netherlands Organisation fo Scientific Research (NWO) under contract
number 614-51-002, and by the Netherlands Research-school for
Astronomy (NOVA). This research has made use of data obtained through
the High Energy Astrophysics Science Archive Research Center Online
Service, provided by the NASA/Goddard Space Flight Center.

\newpage





\end{document}